\documentstyle[preprint,aps]{revtex}
\begin{document}
\draft
\title{Remark on the vectorlike nature of the electromagnetism and the 
electric charge 
quantization }
\author{ C. A de S. Pires}
\address{\tightenlines{Instituto de  F\'{\i}sica Te\'{o}rica, Universidade 
Estadual
Paulista, Rua Pamplona 145, 01405-900 S\~{a}o Paulo, S\~{a}o Paulo, 
Brazil.}}
\maketitle
\begin{abstract}
In this work we study the structure of the electromagnetic interactions and the 
electric charge quantization in gauge theories of electroweak interactions based 
on semi-simple groups.  We show that in the standard model of the electroweak 
interactions the structure of the electromagnetic 
interactions is strongly correlated to the quantization pattern of the electric 
charges. We 
examine these two questions also in all possible chiral bilepton gauge models of 
the electroweak 
interactions. In all they we can explain the vectorlike nature of the 
electromagnetic 
interactions and the 
electric charge quantization together demanding nonvanishing fermion masses and 
the anomaly 
cancellations.  
\end{abstract}
\pacs{PACS numbers: 12.60.cn; 12.90.+b}

Why nature arranges the things so that the electromagnetic interactions among 
fermions are vectorial and  electric charge quantization (ECQ) comes with the 
pattern : $Q_\nu =0$, $Q_e =-e$, $Q_u = \frac{2}{3}e$, $Q_d 
=-\frac{1}{3}e$ is a question  not altogether closed. The QED, the natural place 
to investigate these questions, is unable to explain them because of the 
arbitrariness of 
the $U(1)_{em}$ quantum numbers. Gauge theories of 
electroweak interactions based on semi-simple group face also the same 
difficulty as in QED: 
the 
$U(1)$ factor compounding the semi-simple groups.  

In spite of this difficulty 
the ECQ was analyzed in gauge theories of electroweak 
interactions. 
It was found that in some models 
difficulties may be overcomed by two types of constraints. One type comes from 
the 
requiriment that some fermions be massives. In such models the fermions obtain 
masses through the Yukawa sector. Demanding this sector to be invariant under 
the gauge symmetry  
we get constraints among the $U(1)$ quantum numbers. These are the 
nonvanishing 
fermion masses constraints also called classical constraints. Another type of 
constraints come from the requiriment of theoretical consistency of the model 
which means the model be free from anomalies. Anomaly cancellations\cite{1} are 
expressed in term of  relations among the $U(1)$ quantum numbers. These are the 
quantum constraints.  Using these two constraints the ECQ can be explained in 
some extension of the standard model(SM)\cite{2,3,4,5,6,7,8}. In this work we 
extend such analysis  to include the structure of the 
electromagnetic interactions. Such study will be done in the SM with massless 
and massive neutrinos and in all possibles chiral bilepton gauge(CBGM) versions.

This paper is organized as follows. In Sec. II we analyze the VLNE and the ECQ 
in the SM 
with one and three generations. In sec. III we extend our analysis to the case 
of CBGM. In Sec. IV we 
summarize our 
conclusions. 
 
\section{The electric charge quantization and the vectorlike nature of the 
electromagnetism in the framework of the standard model}
\subsection{ The case of one generation}
In the SM with one generation the quarks and leptons come in the following 
representations
\begin{eqnarray}
&&L_L=
\left (
\begin{array}{c}
\nu \\
e
\end{array}
\right )_L \,\, \sim (\mbox{{\bf 1,2}},Y_L),\,\,\,\,\,\,e_R \sim (\mbox{{\bf 1, 
1}},Y_l),\nonumber \\
&&Q_L=
\left (
\begin{array}{c}
u \\
d
\end{array}
\right )_L \,\, \sim (\mbox{{\bf 3}}, \mbox{{\bf 2}},Y_q),\,\,\,\, u_R \sim 
(\mbox{{\bf 3,1}},Y_u),\,\,\,\, d_R \sim (\mbox{{\bf 3,1}},Y_d) .
\label{1} 
\end{eqnarray}

In order to break symmetry spontaneously and give masses to the gauge bosons 
$W^{\pm}$ and $Z^0$ we need to introduce a Higgs doublet  $\phi \sim 
(\mbox{{\bf 1,2}},Y_\phi)$ that acquires a vacuum expectation value 
\begin{equation}
\langle \phi \rangle_0 \sim
\left (
\begin{array}{c}
0 \\
v 
\end{array}
\right ) .
\label{2} 
\end{equation}

After the spontaneous symmetry breaking(SSB) and due to the mixing among $W^3$ 
and 
$B$ 
we find the following charge operator\cite{4}
\begin{equation}
Q/a=T_3+b/a\frac{Y}{2}.
\label{3}
\end{equation}
Where $a=g/e\sin\theta_W$ and $b=g^{\prime}/e\cos\theta_W$.
Since we want the generator $Q$ unbroken, $Q\langle \phi \rangle_0$ must be 
zero. With this condition we find $a=bY_\phi$.  Using the freedom in assigning 
the scale of the electric charge we set $a=1$\cite{2,4}. Then we have 
$g\sin\theta_W=e$   and $g^{\prime}\cos\theta_W=e/Y_\phi$. 

After these 
steps we find the following electromagnetic interactions among 
the fermions
\begin{eqnarray}
{\cal L}_{em}&=&-\frac{e}{4Y_\phi}[2(Y_L+Y_\phi)\bar \nu_L \gamma^\mu \nu_L 
+\nonumber \\
&&+\bar e \left((-Y_L-Y_l+Y_\phi)+(-Y_L+Y_l+Y_\phi)\gamma_5\right)\gamma^\mu e 
\nonumber 
\\
&&+\bar u \left((-Y_q-Y_u-Y_\phi)+(-Y_q+Y_u-Y_\phi)\gamma_5\right)\gamma^\mu u 
\nonumber 
\\
&&+\bar d \left((-Y_q-Y_d+Y_\phi)+(-Y_q+Y_d+Y_\phi)\gamma_5\right)\gamma^\mu 
d]A_\mu.
\label{4}
\end{eqnarray}
In order to give to  fermions their masses we introduce 
the Yukawa interaction
\begin{equation}
-{\cal L}^Y = g^l \bar L_L \phi e_R + g^u \bar Q_L \stackrel\sim\phi u_R + g^d 
\bar Q_L \phi d_R + {\rm H.c}.
\label{5}
\end{equation}
By demanding ${\rm U}(1)_Y$ gauge invariance we 
find from (\ref{5})
\begin{equation}
Y_L-Y_l -Y_\phi=0,\,\,\,\,\, Y_q -Y_u+Y_\phi=0,\,\,\,\,\, Y_q-Y_d -Y_\phi=0.
\label{6}
\end{equation}
Substituting (\ref{6}) in (\ref{4}) we find the following structure to the 
electromagnetic interactions
\begin{eqnarray}
{\cal L}_{em}&=&-\frac{e}{2Y_\phi}[(Y_L+Y_\phi)\bar \nu_L \gamma^\mu \nu_L 
+\bar e (-Y_L+Y_\phi)\gamma^\mu e\nonumber \\
&&-(Y_q+Y_\phi)\bar u\gamma^\mu u +(-Y_q+Y_\phi)\bar d \gamma^\mu d]A_\mu 
.
\label{7}
\end{eqnarray}

Now we must impose the anomaly cancellation. The 
conditions in (\ref{6}) leaves only two nontrivial anomalies, which are 
sufficient to fix the hypercharges $Y_L$ and $Y_q$ 
\begin{eqnarray}
&&[{\rm SU}(2)_L]^2 {\rm U}(1)_Y \Longrightarrow Y_q=-\frac{1}{3}Y_L,\nonumber 
\\
&&[{\rm U}(1)_Y]^3_Y \Longrightarrow Y_L=-Y_\phi \Longrightarrow 
Y_q=\frac{1}{3}Y_\phi.
\label{8}
\end{eqnarray}
Substituting (\ref{8}) in (\ref{7}) we obtain the 
ECQ 
and the VLNE
\begin{eqnarray}
{\cal L}_{em}=e\bar e\gamma^\mu e A_\mu-\frac{2e}{3}\bar u \gamma^\mu u 
A_\mu+\frac{e}{3}\bar d \gamma^\mu d A_\mu .
\label{9}
\end{eqnarray}

Cancellation of anomalies and demanding the fermions are massives one 
obtains the ECQ and the VLNE in the SM with one generation and massless 
neutrinos.

Adding a right-handed Dirac neutrino, $\nu_R 
\sim (\mbox{{\bf 1,2}},Y_\nu)$, to the SM we get the following electromagnetic 
interaction
\begin{equation}
-\frac{e}{4Y_\phi}\bar \nu_l 
\left((Y_L+Y_\nu+Y_\phi)+(Y_L-Y_\nu+Y_\phi)\gamma_5\right)\gamma^\mu \nu_l 
A_\mu.
\label{10}
\end{equation}
Its Yukawa term is $\bar L_L \tilde \phi \nu_R$. Its $U(1)_Y$ gauge invariance 
provides the 
following relation: $-Y_L+Y_\nu -Y_\phi=0$, which cancels the axial term in 
(\ref{10}). But now  only one nontrivial anomaly constraint remains: 
$[{\rm SU}(2)_L]^2{\rm U}(1)_Y$. Then we have two arbitrary hypercharges from 
the gauge invariance of the Yukawa sector and one constraint from the anomaly 
cancellation. In this case we have no ECQ. This result is the 
dequantization effect\cite{2,3,4,5,6,7}. Nevertheless we have the VLNE 
automatically 
\begin{eqnarray}
{\cal L}_{em}&=&-\frac{e}{2Y_\phi}[(Y_L+Y_\phi)\bar \nu \gamma^\mu \nu 
+(-Y_L+Y_\phi)\bar e \gamma^\mu e \nonumber \\
&&+(\frac{Y_L}{3}+Y_\phi)\bar u\gamma^\mu u + (-\frac{Y_L}{3}+Y_\phi)\bar 
d\gamma^\mu d]A_\mu .
\label{11}
\end{eqnarray}
Babu and Mohapatra made the important 
observation in  Ref. \cite{5} that if 
we suppose that neutrino is a Majorana particle we can fix all the 
hypercharges 
restoring 
thus the ECQ and consequently the VLNE .

\subsection{ The case of three Generations}
With three generations the representation content is
\begin{eqnarray}
&&L_{a_L}=
\left (
\begin{array}{c}
\nu_a \\
e_a
\end{array}
\right )_L  \sim (\mbox{{\bf 1,2}},Y_{L_a}),\,\,\,\,\,\,e_{a_R} \sim (\mbox{{\bf 
1, 
1}},Y_{l_a}),\nonumber \\
&&Q_{a_L}=
\left (
\begin{array}{c}
u_a \\
d_a
\end{array}
\right )_L  \sim (\mbox{{\bf 3,2}},Y_{q_a}),\,\,\,\, u_{a_R} \sim (\mbox{{\bf 
3,1}},Y_{u_a}),\,\,\,\, d_{a_R} \sim (\mbox{{\bf 3,1}},Y_{d_a}) .
\label{12} 
\end{eqnarray}
With $a=1,2,3$ being the flavor index. Now we have the following structure for 
the electromagnetic interactions
\begin{eqnarray}
{\cal L}_{em}&=&-\frac{e}{4Y_\phi}\sum_{a}^{3}[2(Y_{L_a}+Y_\phi )\bar \nu_{a_L} 
\gamma^\mu \nu_{i_L} \nonumber \\
&&+\bar e_a \left((-Y_{L_a}-Y_{l_a}+Y_\phi )+(-Y_{L_a}+Y_{l_a}+Y_\phi )\gamma_5 
\right)\gamma^\mu 
e_a \nonumber \\
&&+\bar u_a 
\left((-Y_{q_a}-Y_{u_a}-Y_\phi)+(-Y_{q_a}+Y_{u_a}-Y_\phi )\gamma_5 
\right)\gamma^\mu u_a 
\nonumber \\
&&+\bar d_a 
\left((-Y_{q_a}-Y_{d_a}+Y_\phi )+(-Y_{q_a}+Y_{d_a}+Y_\phi)\gamma_5 
\right)\gamma^\mu d_a 
]A_\mu.
\label{13}
\end{eqnarray}

The Yukawa sector  takes the following form\cite{4}
\begin{eqnarray}
-{\cal L}^Y = \sum_{a,b}^{1,2,3}[g^l_{aa}\bar L_{a_L} \phi e_{a_R} + g^d_{ab} 
\bar Q_{a_L} 
\phi 
d_{b_R} 
+ g^u_{ab}\bar Q_{a_L} \tilde \phi u_{b_R}] + {\rm H.c} .
\label{14}
\end{eqnarray}

From its  ${\rm U}(1)_Y$ gauge 
invariance we find the following relations among the hypercharges
\begin{equation}
Y_{L_a}-Y_{l_a}-Y_\phi=0,\,\,\,\,\,Y_{q_a}-Y_{u_b}-Y_\phi=0,\,\,\,\,\,Y_{q_a}-Y_
{d_b}+Y_\phi=0,
\label{15}
\end{equation}
From the two last terms above 
we obtain
\begin{equation}
Y_{q_a}=Y_q,\,\,\,\,\,Y_{u_a}=Y_u,\,\,\,\,\,Y_{d_a}=Y_d,
\label{16}
\end{equation}
which lead (\ref{15}) to
\begin{equation}
Y_{L_a}-Y_{l_a}-Y_\phi=0,\,\,\,Y_{q}-Y_{u}-Y_\phi=0,\,\,\,Y_{q}-Y_{d}+Y_\phi=0.
\label{17}
\end{equation}
Substituting (\ref{17}) into (\ref{13}) we obtain the following electromagnetics 
interaction among fermions 
\begin{eqnarray}
{\cal L}_{em}&=&-\frac{e}{2Y_\phi}\sum_{a}^{3}[(Y_{L_a}+Y_\phi)\bar \nu_{a_L} 
\gamma^\mu \nu_{a_L} A_\mu+\bar e_a(-Y_{L_a}+Y_\phi)\gamma^\mu e_a 
A_\mu\nonumber \\ &&-\bar u_a (Y_q+Y_\phi)\gamma^\mu u_a A_\mu+\bar d_a 
(-Y_q+Y_\phi)\gamma^\mu d_a A_\mu .
\label{18}
\end{eqnarray}

After this we have only two nontrivial anomaly constraints  
\begin{eqnarray}
&&[{\rm SU}(2)_L]^2 {\rm U}(1)_Y \Longrightarrow 9Y_q+\sum_{a}^{3}Y_{L_a}=0 
,\nonumber \\
&&[{\rm U}(1)_Y]^3_Y \Longrightarrow 
18Y^3_q-9Y^3_u-9Y^3_d+\sum_{a}^{3}(2Y^3_{L_a}-Y^3_{l_a})=0. 
\label{19}
\end{eqnarray}
These two constraints are insufficient for fix the four arbitrary hypercharges 
in 
(\ref{18}). 
Differently of the SM 
with one generation we have neither explanation to the ECQ 
nor to the VLNE. This is the 
dequantization 
effect\cite{2,3,4,5,6,7}. Nevertheless we have through (\ref{18})  a correlation 
among the quantization pattern and the structure of the electromagnetic 
interactions. Such correlation permits us to conclude that nature arranges the 
things so 
that the  electric 
charge quantization comes with the quantization pattern  $Q_\nu 
=0$, $Q_e =-e$, $Q_u = \frac{2}{3}e$, $Q_d =-\frac{1}{3}e$ because the nature of 
the 
electromagnetic interactions is vectorial. Thus whether we wish explain the ECQ 
we need to take as constraints the nonvanishing fermion masses, anomaly 
cancellations and the VLNE.

Let us suppose  a Dirac-like massive neutrinos. In this case also we have the 
dequantization effect\cite{2,3,4}. Nevertheless the VLNE is automatic
\begin{eqnarray}
{\cal L}_{em}&=&-\frac{e}{2Y_\phi}\sum_{a}^{3}[(Y_L+Y_\phi)\bar \nu_a \gamma^\mu 
\nu_a A_\mu+\bar e_a(-Y_L+Y_\phi)\gamma^\mu e_a A_\mu\nonumber \\ &&-\bar u_a 
(-\frac{Y_L}{3}+Y_\phi)\gamma^\mu u_a A_\mu+\bar d_a 
(-\frac{Y_L}{3}+Y_\phi)\gamma^\mu d_a A_\mu ,
\label{20}
\end{eqnarray}
where $Y_{L_1}=Y_{L_2}=Y_{L_3}=Y_L$\cite{4}. 

If we suppose a Majorana-like massive neutrinos we must restore the 
ECQ and the VLNE like in the case of the SM with one generation\cite{2,3}.

\section{The electric charge quantization and the vectorlike nature of the 
electromagnetism in  chiral bilepton gauge models}

Chiral bilepton gauge models are extensions of the $SU(3)_C\otimes 
SU(2)_L\otimes U(1)_Y$ symmetry group to the $ G_{3X1} = SU(3)_C\otimes 
SU(X)_L\otimes U(1)_N 
$ one, with $X=3,4$. $X=3$ gives us the simplest versions and  $X=4$ 
the largest version. Whether we do not consider exotic leptons  we can have only 
two 
independent  simplest versions of CBGM\cite{9,10} and one largest 
version\cite{11}. Their key feature are bilepton gauge bosons with lepton 
number $L=\pm2$ and exotic quarks. They give in some sense a answer 
to the family problem\cite{12} because they require a minimal of three families 
to cancel anomalies. They are a multi-higgs model, nevertheless  FCNC 
with the standard neutral gauge boson is strongly suppressed\cite{13}. Some 
studies of their 
phenomenology were done in \cite{14}.
 
\subsection{Version A}
 
Here we analyze the VLNE and the ECQ in a CBGM based in the ${\rm SU}(3)_C 
\otimes {\rm SU}(3)_L \otimes {\rm U}(1)_N$ symmetry which has as electric 
charge operator the following linear combinations of their diagonal 
generators\cite{9}
\begin{eqnarray}
Q=\frac{1}{2}(\lambda _3 -\sqrt{3}\lambda_8)+bN  , 
\label{21}
\end{eqnarray}
where $N$ is the generator operator of the  ${\rm U}(1)_N$ group  $\lambda_3$  
and  $\lambda_8$ the two diagonal Gell-mann matrices.
 
The mininal set of scalars necessary to give correct masses to the fermions and 
bosons are three triplets and one sextet. They get  vacuum expectation value 
different from zero and transform by ${\rm SU}(3)_C \otimes {\rm SU}(3)_L 
\otimes {\rm U}(1)_N$ in the following manner 
\begin{eqnarray}
&&\langle \eta \rangle_0 = \left( 
\begin{array}{c}
v_\eta \\ 
0 \\ 
0
\end{array}
\right)\sim (\mbox{{\bf 1,3}},N_\eta ) \,\,\,,\,\,\,\langle \rho \rangle_0 
=\left( 
\begin{array}{c}
0 \\ 
v_\rho \\ 
0
\end{array}
\right)\sim (\mbox{{\bf1,3}},
N_\rho )  \nonumber \\
&&\langle \chi \rangle_0 = \left( 
\begin{array}{c}
0 \\ 
0 \\ 
v_\chi
\end{array}
\right)\sim (\mbox{{\bf 1,3}},N_\chi ) , \langle S \rangle_0 = \left( 
\begin{array}{lcr}
0 & 0 & 0 \\ 
0 & 0 & v_{\sigma_2} \\ 
0 & v_{\sigma_2} & 0
\end{array}
\right)\sim (\mbox{{\bf 1,3}},N_S) .  
\label{22}
\end{eqnarray}

Requiring the electric charge operator annihilates the vacuum of the scalars we 
set 
\begin{equation}
 N_\eta =0,\,\,\,  b=\frac{1}{N_\rho },\,\,\,N_\chi =-N_\rho 
,\,\,\,N_S=0.  
\label{23}
\end{equation}

After these steps the electric charge operator acquire the following form
\begin{equation}
Q=\frac{1}{2}(\lambda _3 -\sqrt{3}\lambda_8)+\frac{N}{N_\rho}.  
\label{24}
\end{equation}

The fermions come in the following representations
\begin{eqnarray}
&&L_{a_L}=
\left (
\begin{array}{c}
\nu_a \\
e_a \\
e_a^c
\end{array}
\right )_L \,\, \sim (\mbox{{\bf 1,3}},N_{L_a}),\nonumber \\
&&Q_{1_L}=
\left (
\begin{array}{c}
u_1 \\
d_1 \\
J_1
\end{array}
\right )_L \,\, \sim (\mbox{{\bf 3,3}},N_{Q_1}),\nonumber \\
&&u_{1R} \sim (\mbox{{\bf 3, 1}},N_{u_1})\,\,\,\,\,\,d_{1_R} \sim 
(\mbox{{\bf 
3,1}},N_{d_1})\,\,\,\,\,\, J_{R_1} \sim (\mbox{{\bf 3,1}},N_{J_1}),\nonumber \\
&&Q_{i_L}=
\left (
\begin{array}{c}
d_i \\
-u_i \\
J_i
\end{array}
\right )_L \,\, \sim (\mbox{{\bf 3}}, \mbox{{\bf 
3}}^*,N_{Q_i}),\nonumber \\
&&d_{i_R} \sim (\mbox{{\bf 3,1}},N_{d_i}),\,\,\,\,\,\, u_{i_R} \sim 
(\mbox{{\bf 
3,1}},N_{u_i}),\,\,\,\,\,\, J_{i_R} \sim (\mbox{{\bf 3,1}},N_{J_i}) ,
\label{25} 
\end{eqnarray}
with $a=1,2,3$ and $i=2,3$ being  flavor index.  There are no lepton singlets. 
The quarks $u$'s  e  $d$'s are the usual ones 
with $J$'s 
being 
the exotic quarks.

After SSB ${\rm SU}(3)_C \otimes {\rm SU}(3)_L 
\otimes 
{\rm U}(1)_N \rightarrow {\rm SU}(3)_C \otimes {\rm U}(1)_{em}$ we find the 
following structure for the electromagnetic interactions
\begin{eqnarray}
{\cal L}_{em}&=&-\frac{e}{2N_\rho}\sum_{a}^{1,2,3}\sum_{i}^{2,3}[2N_{L_a}\bar 
\nu_{a_L}\gamma^\mu \nu_{a_L} +2\bar e_a \left(-N_\rho + 
N_{L_a}\gamma_5\right)\gamma^\mu e_a  \nonumber \\
&&+\bar u_1\left((N_{Q_1} + N_{u_1}) - (-N_{Q_1} + 
N_{u_1})\gamma_5\right)\gamma^\mu u_1  \nonumber \\
&&+\bar u_i\left((N_{Q_i} + N_{u_i} + N_\rho) - (N_{Q_i} - N_{u_i} + 
N_\rho)\gamma_5\right)\gamma^\mu u_i \nonumber \\
&&+\bar d_1\left((N_{Q_1} + N_{d_1} -N_\rho) - (N_{Q_1} - 
N_{d_1}-N_\rho)\gamma_5\right)\gamma^\mu d_1  \nonumber \\
&&+\bar d_i\left((N_{Q_i} + N_{d_i}) - (N_{Q_i} - 
N_{d_i})\gamma_5\right)\gamma^\mu d_i \nonumber \\
&&+\bar J_1\left((N_{Q_1} + N_{J_1}+N_\rho) - (N_{Q_1} + 
N_{J_1}-N_\rho)\gamma_5\right)\gamma^\mu J_1  \nonumber \\
&&+\bar J_i\left((N_{Q_i} + N_{J_i}-N_\rho) - (N_{Q_i} - 
N_{J_i}-N_\rho)\gamma_5\right)\gamma^\mu J_i]A_\mu.
\label{26}
\end{eqnarray}

From the ${\rm U}(1)_N$ invariance of the Yukawa sector\cite{15}
\begin{eqnarray}
-{\cal L}_{Y} &=&\frac{1}{2}G_{ab }\bar{L_{aL }^c}S^* L_{bL}  \nonumber \\
&+&\lambda _1\bar{Q}_{1L}J_{1R}\chi +\lambda _{ij}\bar{Q}_{iL}J_{jR}\chi 
^{*}
\nonumber \\
&+&\lambda^{\prime} _{1a}\bar{Q}_{1L}d_{aR}\rho +\lambda^{\prime} _{ia}\bar{Q}
_{iL}u_{aR}\rho^*  \nonumber \\
&+&\lambda^{\prime \prime} _{1a}\bar{Q}_{1L}u_{aR}\eta +\lambda^{\prime \prime} 
_{ia}\bar{Q}
_{iL}d_{aR}\eta ^{*}\,\,\,+\,\,\,H.c., 
 \label{27}
\end{eqnarray}
we find the following relations between the $N$ quantum numbers\cite{8} 
\begin{eqnarray}
&&N_{u_1}=N_{u_2}=N_{u_3}=N_u,\nonumber \\
&&N_{d_1}=N_{d_2}=N_{d_3}=N_d,  \nonumber \\
&&N_{Q_2}=N_{Q_3}=N_Q,  \nonumber \\
&&N_{J_2}=N_{J_3}=N_J,
\label{28}
\end{eqnarray}
and
\begin{eqnarray}
&&N_{L_1}=N_{L_2}=N_{L_3}=0,\nonumber \\
&&N_J=N_Q-N_\rho ,  \nonumber \\
&&N_d=N_Q,  \nonumber \\
&&N_u=N_Q+N_\rho ,  \nonumber \\
&&N_{J_1}=N_Q+2N_\rho ,  \nonumber \\
&&N_{Q_1}=N_Q+N_\rho.
\label{29}
\end{eqnarray}

Substituting these relations in (\ref{26}) we obtain the following 
electromagnetic interactions
\begin{eqnarray}
{\cal L}_{em}&=& e \bar e_a \gamma^\mu e_a A_\mu  \nonumber \\
&&-\frac{e}{N_\rho}[(N_Q + N_\rho)\bar u_a \gamma^\mu u_a +N_Q \bar d_a 
\gamma^\mu d_a  \nonumber \\
&&+(N_Q + 2N_\rho)\bar J_1\gamma^\mu J_1  +(N_Q-N_\rho)\bar J_i \gamma^\mu J_i] 
A_\mu.
\label{30}
\end{eqnarray}
Note that the classical constraints alone provide the VLNE and the ECQ among the 
leptons.

After the relations in (\ref{29}) only one nontrivial anomaly 
constraint remains
\begin{eqnarray}
[SU(3)_L]^2U(1)_N\Longrightarrow 
3N_{Q_1}+3N_{Q_2}+3N_{Q_3}+N_{L_1}+N_{L_2}+N_{L_3}=0,  
\label{31}
\end{eqnarray}
which together with the relations in (\ref{29}) gives $N_Q =-N_\rho /3$. This 
fixes uniquely the $N$'s quantum numbers as a function of $N_\rho$, furnishing 
thereby the electric charge quantization of all fermions and the VLNE
\begin{eqnarray}
{\cal L}_{em}&=&e\bar e_a \gamma^\mu e_a A_\mu  +\nonumber \\
&&-\frac{2e}{3}\bar u_a\gamma^\mu u_a A_\mu  +\frac{e}{3}\bar d_a \gamma^\mu d_a 
A_\mu +\nonumber \\
&&-\frac{5e}{3}\bar J_1\gamma^\mu J_1 A_\mu  +\frac{4e}{3}\bar J_i \gamma^\mu 
J_i A_\mu.
\label{32}
\end{eqnarray}

Let us now analyze this model with massive neutrinos. In the case of Majorana 
neutrino, it is sufficient to modify the sextet of scalars to
\begin{eqnarray}
\langle S \rangle_0 = \left( 
\begin{array}{lcr}
v_{\sigma_1} & 0 & 0 \\ 
0 & 0 & v_{\sigma_2} \\ 
0 & v_{\sigma_2} & 0
\end{array}
\right)\sim (\mbox{{\bf 1,3}},N_S ) .  
\label{33}
\end{eqnarray}
This does not change the above steps. So the result in (\ref{32}) remains the 
same 
with or without Majorana neutrinos.

If we add right-handed neutrinos, $\nu_{a_R} \sim (\mbox{{\bf 1,1}},N_{R_a} )$, 
their electromagnetic 
interaction  become this
\begin{equation}
{\cal L}_{em}^{\nu}=-\frac{e}{N_\rho}\sum_{a}^{3}(N_{L_a}+N_{R_a})\bar \nu_a 
\gamma^\mu 
\nu_a A^\mu.
\label{34}
\end{equation}	
From their Yukawa 
interactions $\bar L_{a_L} \eta \nu_{a_R}$, we find : $N_{R_a}-N_{L_a}=0$, which 
together with (\ref{29}) gives  $N_{R_a} = 0$, annulling 
their electromagnetic interactions. In short, in this simplest CBGM 
version we explain, with or without neutrinos are massives, the ECQ and the 
VLNE.

\subsection{Version B}

This is another possible variant of the simplest CBGM versions. Its  Higgs 
sector is more 
economic than the one in the first version. It present Dirac-like massive 
neutrinos inevitably in tree level. Its fermion content is the following

\begin{eqnarray}
&&L_{a_L}=
\left (
\begin{array}{c}
\nu_a \\
e_a \\
\nu_a^c
\end{array}
\right )_L \,\, \sim (\mbox{{\bf 1,3}},N_{L_a}),\,\,\,e_{a_R}\sim (\mbox{{\bf 
1,1}},N_{R_a}),\nonumber \\
&&Q_{\alpha_L}=
\left (
\begin{array}{c}
d_\alpha \\
-u_\alpha \\
J_\alpha
\end{array}
\right )_L \,\, \sim (\mbox{{\bf 3}}, \mbox{{\bf 
3}}^*,N_{Q_\alpha}),\nonumber \\
&&u_{\alpha_R} \sim (\mbox{{\bf 3, 1}},N_{u_\alpha})\,\,\,\,\,\,d_{\alpha_R} 
\sim 
(\mbox{{\bf 
3,1}},N_{d_\alpha})\,\,\,\,\,\, J_{R_\alpha} \sim (\mbox{{\bf 
3,1}},N_{J_\alpha}), \nonumber \\
&&Q_{3L}=
\left (
\begin{array}{c}
u_3 \\
d_3 \\
J_3
\end{array}
\right )_L \,\, \sim(\mbox{{\bf 3,3}},N_{Q_3}) ,\nonumber \\
&&d_{3_R} \sim (\mbox{{\bf 3,1}},N_{d_3}),\,\,\,\,\,\, u_{3_R} \sim 
(\mbox{{\bf 
3,1}},N_{u_3}),\,\,\,\,\,\, J_{3R} \sim (\mbox{{\bf 3,1}},N_{J_3}),
\label{35} 
\end{eqnarray}
with $a=1,2,3$  and $\alpha=1,2$ being the flavor index.

Its  electric charge operator takes the following linear combination\cite{10}
\begin{eqnarray}
Q=\frac{1}{2}(\lambda _3 -\frac{1}{\sqrt{3}}\lambda_8)+b^{\prime}N . 
\label{36}
\end{eqnarray}

We need three triplets of scalars to break spontaneouly the symmetry  and give 
mass to the fermions. Their  vacuum expectation value is different from zero and 
transforms by ${\rm SU}(3)_C \otimes {\rm SU}(3)_L \otimes {\rm U}(1)_N$ in the 
following manner 
\begin{eqnarray}
&&\langle \eta \rangle_0 = \left( 
\begin{array}{c}
v_\eta \\ 
0 \\ 
0
\end{array}
\right)\sim (\mbox{{\bf 1,3}},N_\eta ) \,\,\,,\,\,\,\langle \rho \rangle_0 
=\left( 
\begin{array}{c}
0 \\ 
v_\rho \\ 
0
\end{array}
\right)\sim (\mbox{{\bf1,3}},
N_\rho )  \nonumber \\
&&\langle \chi \rangle_0 = \left( 
\begin{array}{c}
0 \\ 
0 \\ 
v_\chi
\end{array}
\right)\sim (\mbox{{\bf 1,3}},N_\chi ) .  
\label{37}
\end{eqnarray}

With the requiriment that the electric charge operator annihilates the vacuum of 
the scalars we set 
\begin{equation}
 N_\rho =-2N_\eta ,\,\,\,N_\chi =N_\eta 
,\,\,\,b^{\prime}=-\frac{1}{3N_\eta}.  
\label{38}
\end{equation}

After the break of the symmetry ${\rm SU}(3)_C \otimes {\rm SU}(3)_L \otimes 
{\rm U}(1)_N \rightarrow {\rm SU}(3)_C \otimes {\rm U}(1)_{em}$ we find the 
following structure for their electromagnetic interactions
\begin{eqnarray}
{\cal 
L}_{em}&=&-\frac{e}{6N_\eta}\sum_{a}^{1,2,3}\sum_{\alpha}^{1,2}[(N_{L_a}-N_\eta)
\bar \nu_a\gamma_5 \gamma^\mu \nu_a \nonumber \\
&& \bar e_a \left((-N_{L_a}-N_{R_a}-2N_\eta) 
-(-N_{L_a}+N_{R_a}-2N_\eta)\gamma_5\right)\gamma^\mu e_a  \nonumber \\
&&+\bar u_{\alpha}\left((-N_{Q_\alpha}-N_{u_\alpha}+2N_\eta) - (-N_{Q_\alpha} + 
N_{u_\alpha}+2N_\eta)\gamma_5\right)\gamma^\mu u_\alpha  \nonumber \\
&&+\bar u_3\left((-N_{Q_3} - N_{u_3} + N_\eta) - (-N_{Q_3} + N_{u_3} + 
N_\eta)\gamma_5\right)\gamma^\mu u_3 \nonumber \\
&&+\bar d_{\alpha}\left((-N_{Q_\alpha} - N_{d_\alpha} -N_\eta) - (-N_{Q_\alpha} 
+ N_{d_\alpha}-N_\eta)\gamma_5\right)\gamma^\mu d_\alpha  \nonumber \\
&&+\bar d_3\left((-N_{Q_3} - N_{d_3}-2N_\eta) - (-N_{Q_3} + 
N_{d_3}-2N_\eta)\gamma_5\right)\gamma^\mu d_3 \nonumber \\
&&+\bar J_\alpha\left((-N_{Q_\alpha} - N_{J_\alpha}-N_\eta) - (-N_{Q_\alpha} + 
N_{J_\alpha}-N_\eta)\gamma_5\right)\gamma^\mu J_\alpha  \nonumber \\
&&+\bar J_3\left((-N_{Q_3} - N_{J_3}+N_\eta) - (-N_{Q_3} + 
N_{J_3}+N_\eta)\gamma_5\right)\gamma^\mu J_3]A_\mu.
\label{39}
\end{eqnarray}

The Yukawa sector here is\cite{10}
\begin{eqnarray}
-{\cal L}_{Y} &=&G_{ab }\epsilon^{lmn}(\bar L_{a_L })_l( L_{b_L})_m(\rho^*)_n 
+G^{\prime}_{ab }\bar L_{a_L }e_{b_R}\rho \nonumber \\
&+&\lambda _1\bar{Q}_{3_L}J_{3_R}\chi +\lambda _{2\alpha 
\beta}\bar{Q}_{\alpha_L}J_{\beta_R}\chi 
^{*}
\nonumber \\
&+&\lambda _{1a}\bar{Q}_{3_L}d_{a_R}\rho +\lambda_{2\alpha a}\bar{Q}_{\alpha 
L}u_{a_R}\rho^*  \nonumber \\
&+&\lambda_{3a}\bar{Q}_{3_L}u_{a_R}\eta +\lambda_{4\alpha a}\bar{Q}
_{\alpha_L}d_{a_R}\eta ^{*}\,\,\,+\,\,\,H.c., 
 \label{40}
\end{eqnarray}
with $a,b=1,2,3$ and $\alpha,\beta =1,2$. From ${\rm U}(1)_N$ invariance, this 
sector supplies us with the following relations between the $N$ quantum numbers 
\begin{eqnarray}
&&N_{u_1}=N_{u_2}=N_{u_3}=N_u,\nonumber \\
&&N_{d_1}=N_{d_2}=N_{d_3}=N_d,  \nonumber \\
&&N_{Q_1}=N_{Q_2}=N_Q,  \nonumber \\
&&N_{J_1}=N_{J_2}=N_J,
\label{41}
\end{eqnarray}
and
\begin{eqnarray}
&&N_{L_1}=N_{L_2}=N_{L_3}=N_\eta,\nonumber \\
&&N_{R_1}=N_{R_2}=N_{R_3}=3N_\eta,\nonumber \\
&&N_u=N_Q - 2N_\eta ,  \nonumber \\
&&N_d=N_Q +N_\eta,  \nonumber \\
&&N_{Q_3}=N_Q -N_\eta ,  \nonumber \\
&&N_{J_3}=N_Q -2N_\eta ,  \nonumber \\
&&N_J=N_Q +N_\eta.\nonumber \\
\label{42}
\end{eqnarray}
Substituting these relations into (\ref{39}) we obtain
\begin{eqnarray}
{\cal L}_{em}&=&e\bar e_a \gamma^\mu e_a A_\mu \nonumber \\
&&-\frac{e}{3N_\eta}[(-N_Q + 2N_\eta)\bar u_a\gamma^\mu u_a  -(N_Q + N_\eta)\bar 
d_a\gamma^\mu d_a \nonumber \\
&&+(N_Q + 2N_\eta)\bar J_3\gamma^\mu J_3 -(N_Q + N_\eta)\bar J_\alpha \gamma^\mu 
J_\alpha ]A_\mu.
\label{43}
\end{eqnarray}
Note that,as in the version A, the classical constraints lead to  ECQ and the 
VLNE 
in the leptonic 
sector while in the quark sector lead only to the VLNE. 

Again, as in the previous section, only one nontrivial anomaly 
constraint remains
\begin{eqnarray}
[SU(3)_L]^2U(1)_N\Longrightarrow 
3N_{Q_1}+3N_{Q_2}+3N_{Q_3}+N_{L_1}+N_{L_2}+N_{L_3}=0,  
\label{44}
\end{eqnarray}
which together with the relations in (\ref{42}) gives $N_Q = 0$. This  result 
fixes uniquely the $N$'s quantum numbers in function of $N_\eta$ explaining the 
ECQ for all fermions and leading to the VLNE
\begin{eqnarray}
{\cal L}_{em}&=&e\bar e_a \gamma^\mu e_a A_\mu  +\nonumber \\
&&-\frac{2e}{3}\bar u_a\gamma^\mu u_a A_\mu  +\frac{e}{3}\bar d_a \gamma^\mu d_a 
A_\mu +\nonumber \\
&&\frac{e}{3}\bar J_\alpha \gamma^\mu J_\alpha A_\mu  -\frac{2e}{3}\bar J_3 
\gamma^\mu J_3 A_\mu.
\label{45}
\end{eqnarray}

 The largest CBGM version is based on the ${\rm SU}(3)_C \otimes {\rm SU}(4)_L 
\otimes {\rm 
U}(1)_N$ symmetry group\cite{11}. In it the ECQ takes place  in the same way as 
in the first simplest version, as was showed recently in \cite{16}. Thus in it 
the VLNE and the ECQ 
must be explained in the same way as in the simplest 
versions A. We close this section saying that in CBGM inevitably we explain the 
ECQ and the structure of the electromagnetic interactions.    
  
\section{Conclusions}

In this paper we have examined the correlation among ECQ and  VLNE in some gauge 
theory of electroweak interactions in
order to understand the structure of one of the four fundamental forces of the 
nature. Depending on the model and  on their representation content  ECQ and 
the VLNE are strongly correlated. This is the case with the SM with three 
generations and massless neutrinos. In this case, through the classical and 
quantum constraints, we do not explain neither the ECQ nor the VLNE. 
Nevertheless 
we can understand why the electric charge is quantized with the pattern required 
by  
nature through a correlation among the ECQ and the VLNE obtained in (\ref{18}). 
There we can see that 
such required pattern occurs because the QED is vectorial. Also such correlation 
say that whether we wish explain the ECQ we must require as constraints  
the nonvanishing fermion masses, the anomaly cancellations and the VLNE. In the 
case of Dirac-like massive 
neutrinos we lose such correlations in the sense that we have the VLNE but not 
the ECQ. In the case of Majorana-like massive neutrinos we restore the VLNE and 
the ECQ. In  chiral bilepton gauge model we can explain the ECQ and the VLNE 
together.  This takes place in all versions, with or without massive neutrinos, 
through the nonvanishing fermion masses and anomaly cancellations. These results 
make CBGM an interesting extension of the SM. Principally whether  we hope that 
a final theory of matter and forces explains the VLNE and the ECQ together.

\acknowledgements

I thank to J. C. Montero, V. Pleitez and M. Nowakowski by the encouragement and 
critical
suggestions and also thank  M. C. Tijero for reading the manuscript. This 
work 
was supported by the
Coordena\c {c}\~{a}o de Aperfei\c {c}oamento de Pessoal de N\'{\i}vel
Superior (CAPES).


\begin{thebibliography}{99}

\bibitem{1} S. Adler, Phys. Rev. {\bf 177}, 2426 (1969); J. S. Bell, R. Jackiw, 
Nuovo 
Cimento. {\bf A60}, 49 (1969).  


\bibitem{2}   R. Foot, G. C. Joshi, H. Lee and R. R. Volkas, Mod. Phys. Lett. 
{\rm A} {\bf 5}, 2721 (1990);  R. Foot, H. Lee and R.R. Volkas, J. Phys. {\rm G} 
{\bf 19}, 269 (1993).


\bibitem{3}  K. S. Babu and R. N. Mohapatra, Phys. Rev. {\rm D}. {\bf 41}, 271
(1990).

\bibitem{4}  J. Sladkowski and M. Zralek, Phys. Rev. {\rm D}. {\bf 45}, 1701 
(1992).


\bibitem{5}  K. S. Babu and R. N. Mohapatra, Phys. Rev. Lett {\bf 63},
938 (1989).

\bibitem{6}  C. Q. Geng and R.E Marshak, Phys. Rev. {\rm D}. {\bf 39}, 693 
(1989); C. Q. Geng, {\it ibid}. {\bf 41}, 1292 (1992); E.
Golowich and P. B. Pal, Phys. Rev. {\bf 41}, 3757 (1990); S. Rudaz, Phys. Rev. 
{\bf 41}, 2619 (1990); M. Nowakowski, A. Pilaftsis, Phys. Rev. {\rm D}. {\bf 
48}, 259 (1993); N. G. Deshpande, University of Oregon Repot No. OITS-107, 
1979(unpublished).


\bibitem{7}  K. S. Babu and R. N. Mohapatra, Phys. Rev. {\rm D}. {\bf  42}, 
3866 (1990).

\bibitem{8}  C. A. de S. Pires and O. P. Ravinez, Phys. Rev. {\rm D}. {\bf 58}, 
035008  (1998).



\bibitem{9}  F. Pisano and V. Pleitez, Phys. Rev. {\rm D}. {\bf 46}, 410 
(1992); H. 
Frampton, Phys. Rev. Lett. {\bf 69}, 2889 (1992).


\bibitem{10}  J. C. Montero, F. Pisano and V. Pleitez, Phys. Rev. {\rm D}. {\bf  
47}, 
2918 (1993); H. N. Long; Phys. Rev. {\rm D}. {\bf  53},
 
437 (1996).


\bibitem{11}  V. Pleitez, Phys. Rev. {\rm D}. {\bf 53}, 514 (1996).


\bibitem{12} F. Pisano, Mod. Phys. Lett. {\bf A11}, 2639 (1996).

 
\bibitem{13} J. T. Liu, D. Ng, Phys. Rev. {\rm D}. {\bf  
50}, 
548 (1994);


\bibitem{14} J. C. Montero, C. A de S Pires and V. Pleitez, hep-ph/9812306;  
F. Pisano, J. A. Silva-Sobrinho and M. D. Tonasse; Phys. Rev. {\rm D}. {\bf 58}, 
057703 (1998); Prashanta Das, Pankaj Jain, hep - ph/9808256; P. H. Frampton, 
Xiao-hu Guan, Mod. Phys. Lett. {\rm A}{\bf 13}, 2621, (1998); J. C. Montero, F 
V. Pleitez and M. C. Rodrigues, Phys. Rev. {\rm D}. {\bf  
58}, 
097505, (1998); J. C. Montero, V. Pleitez and M. C. Rodrighues, {\it ibid}. {\rm 
D}. {\bf  
58}, 
094026, (1998). 


 
\bibitem{15}  R. Foot, O. F. Hern\'{a}ndez, F. Pisano and V. Pleitez, 
Phys.Rev, 
{\rm D}. {\bf 47},4158, (1993).



\bibitem{16}  A. Doff and F. Pisano, hep-ph/9812303.



\end{thebibliography}
\end{document}